\documentclass[11pt,draft]{article}

\usepackage{amsfonts,amssymb}

\usepackage{a4wide}

\newcommand{\keywords}[1]{\par\noindent \textbf{Keywords:} #1}
\newcommand{\PACS}[1]{\par\noindent\textbf{PACS:} #1}

\newcommand{\Lfour}{\mathbb{L}^4}

\newtheorem{theorem}{Theorem}
\newenvironment{proof}{\par\noindent\textit{Proof:}}{}

\newcommand{\qed}{\hfill$\Box$}

\newtheorem{corollary}{Corollary}
\newtheorem{rema}{Remark}
\newenvironment{remark}{\begin{rema}\rm }{\end{rema}}
\newtheorem{proposition}{Proposition}
\newtheorem{examp}{Example}
\newenvironment{example}{\begin{examp}\rm }{\end{examp}}

\begin{document}

\title{Marginally trapped surfaces in Minkowski 4-space invariant
under a rotation subgroup of the Lorentz group\thanks{The first
author was partially supported by the Research Foundation -
Flanders project G.0432.07. Both authors were partially supported
by the Spanish MEC Grant MTM2007-60731 with FEDER funds and the
Junta de Andaluc\'{\i}a Regional Grant P06-FQM-01951. This work
was started when the first author was a postdoctoral researcher at
the Section of Geometry of the Katholieke Universiteit Leuven.}}

\title{Marginally trapped surfaces in Minkowski 4-space}

\author{Stefan Haesen \\
\small{Simon Stevin Institute for Geometry} \\
\small{Wilhelminaweg 1} \\
\small{2042 Zandvoort} \\
\small{The Netherlands} \\
\small{\texttt{Stefan.Haesen@geometryinstitute.org}} \\ \\
Miguel Ortega \\
\small{Department of Geometry and Topology} \\
\small{Universidad de Granada} \\
\small{18071 Granada, Spain} \\
\small{\texttt{miortega@ugr.es}}}

\date{}

\maketitle

\begin{abstract}
A local classification of spacelike surfaces in Minkowski 4-space,
which are invariant under spacelike rotations, and with mean
curvature vector either vanishing or lightlike, is obtained.
Furthermore, the existence of such surfaces with prescribed
Gaussian curvature is shown. A procedure is presented to glue
several of these surfaces with intermediate parts where the mean
curvature vector field vanishes. In particular, a local
description of marginally trapped surfaces invariant under
spacelike rotations is exhibited. \keywords{Lightlike mean
curvature vector, marginally trapped surfaces, Lorentz group}
\end{abstract}


\section{Introduction}

In General Relativity, a spacelike surface in a 4-dimensional
Lorentzian manifold is called \textit{marginally trapped} if its
mean curvature vector is proportional to one of the null normals,
by an either positive or negative function. When such function is
arbitrary, the surface is called marginally outer trapped, or
MOTS, for short. The study of these families of surfaces has been
quite active in recent years (see for instance \cite{bray,huisken,senovilla1}).

In general, it is customary to ask these surfaces to be closed,
i.~e., compact and without boundary. However, some results
concerning the non-existence of closed MOTS can be found in the
literature. Among the classical ones, a result due to R. Penrose,
\cite{penrose} (see also \cite{mars}), implies the non-existence
of closed MOTS in the Minkowski spacetime when it bounds a compact
domain. Moreover, in \cite{huisken}, it is proved that in any
strictly static spacetime, no marginally trapped surface which is
not globally extremal (i.~e., its mean curvature vector is not
zero at least in a point) can exist. Also, A. Carrasco and M.
Mars, \cite{CM}, have shown the non-existence of MOTS bounding a
domain and entering a region of a static spacetime where the Killing vector field is timelike, 
and with the additional assumptions of dominant energy
condition and an outer untrapped barrier. Thus, some authors are
beginning to relax the definition, letting the surface to be
non-compact.

In order to gain some idea of the properties of marginally trapped
surfaces in particular spacetimes, classification results were
obtained for marginally trapped surfaces with positive relative
nullity in Lorentzian space forms \cite{chen1} and in
Robertson-Walker spaces \cite{chen2}. In \cite{haesenortega}
marginally trapped surfaces which are invariant under a boost
transformation in 4-dimensional Minkowski space were studied.

We consider the Minkowski 4-space $\Lfour$ endowed with its
standard metric $-\mbox{d}x_{1}^{2} + \mbox{d}x_{2}^{2}
+\mbox{d}x_{3}^{2} + \mbox{d}x_{4}^{2}$. In this paper, we are
interested in studying marginally trapped surfaces in Minkowski
4-space which are invariant under the following group of
isometries:
\[
\mathbf{G}_{s}   =  \left\{ B_{\theta} =\left( \begin{array}{cccc}
1 & 0 & 0 & 0 \\
0 & 1 & 0 & 0 \\
0 & 0 & \cos(\theta) & \sin(\theta) \\
0 & 0 & -\sin(\theta) & \cos(\theta)  \end{array} \right) \, : \,
\theta\in\mathbb{R}\right\}.
\]
{\noindent} Due to the previous non-existence results, we relax
the definition of these surfaces, in the sense that we consider
non-closed marginally trapped surfaces (i.~e. either non-compact
or with boundary). However, with little more efforts, it is
possible to study a more general family of surfaces, namely, those
whose mean curvature vector is either lightlike or zero, and
invariant under $\mathbf{G}_s$. Thus, the main result of this
paper is Theorem \ref{th:class1}, where a local classification of
such surfaces is obtained. In particular, this classification
includes marginally trapped surfaces and those surfaces with
vanishing mean curvature vector, which are invariant by
$\mathbf{G}_s$. For the sake of simplicity, we say that a surface
is extremal at a point $p$ if its mean curvature vector field is
zero at $p$. Needless to say, an extremal surface has everywhere
vanishing mean curvature vector field.

Further, a gluing procedure is presented to construct
$\mathbf{G}_{s}$-invariant spacelike surfaces for which the mean
curvature vector is lightlike or zero on certain parts. This
allows to obtain examples of  various surfaces occurring in the
classification given in \cite{senovilla}. The point is that the
examples constructed using this method have up to infinitely many
regions where the mean curvature vector of each such region can be
chosen to be future or past-pointing \textit{as desired}, and
among two consecutive regions, there is an extremal subset.

In the final section, it is shown that it is possible to construct
surfaces which are invariant by $\mathbf{G}_s$, whose mean
curvature vector is lightlike or zero, and with prescribed
Gaussian curvature. In particular, those with constant Gaussian
curvature are given explicitly.

The main mathematical tool consist of the local theory of
surfaces. Its origins go back almost two centuries ago, when
C.~F.~Gauss \cite{gauss} and other authors started its development
for surfaces in the Euclidean 3-space. Since then, this powerful
theory has been used successfully in an overwhelming number of
situations. Nowadays, this is the standard technique to study
surfaces in Mathematics. At the end of the day, it can be
summarized in a small collection of formulae,  bringing to light
interesting geometric properties of surfaces in Physics.

Finally, the authors would like to thank the referees and the
editors for many useful comments, which helped to improve this
paper.


\section{Preliminaries}

Let $(\mathbb{L}^{4},\widetilde{g})$ be the 4-dimensional
Lorentz-Minkowski space with the flat metric given in local
coordinates by

\[ \widetilde{g} = -\mbox{d}x_{1}^{2} + \mbox{d}x_{2}^{2}
+\mbox{d}x_{3}^{2} + \mbox{d}x_{4}^{2}. \]

{\noindent}For a connected immersed surface $S$ in
$\mathbb{L}^{4}$, we denote by $g$ the induced metric on $S$. We
will assume that this metric $g$ is positive-definite, i.e., the
surface is \emph{spacelike}. Let $\widetilde{\nabla}$ and $\nabla$
denote the Levi-Civita connections on
$(\mathbb{L}^{4},\widetilde{g})$ and $(S,g)$, respectively. Then,
if $X$ and $Y$ are two smooth vector fields tangent to $S$, the
Gauss formula gives the decomposition of the vector
$\widetilde{\nabla}_{X}Y$ into its tangential and normal parts,
i.e.,

\[ \widetilde{\nabla}_{X}Y = \nabla_{X}Y + \mathbf{K}(X,Y), \]

{\noindent}where
$\mathbf{K}:\mathfrak{X}(S)\times\mathfrak{X}(S)\rightarrow
\mathfrak{X}^{\perp}(S)$ is called the \emph{shape tensor} or
\emph{second fundamental form} of $S$ in $\mathbb{L}^{4}$. If
$\eta$ is a normal vector to the surface, the Weingarten formula
gives the decomposition of the vector $\widetilde{\nabla}_{X}\eta$
into its tangential and normal parts, i.e.,

\[ \widetilde{\nabla}_{X}\eta = -A_{\eta}(X) +
\nabla^{\perp}_{X}\eta, \]

{\noindent}where $\nabla^{\perp}$ is the normal connection in the
normal bundle of $S$ and the endomorphism $A_{\eta}$ on
$\mathfrak{X}(S)$ is called the \emph{shape operator} associated
with $\eta$. The shape tensor and shape operator are related by
$\widetilde{g}(\mathbf{K}(X,Y),\eta) = g(A_{\eta}(X),Y)$. The
\emph{mean curvature vector} $\mathbf{H}$ is defined as the
normalized trace of the shape tensor,

\[ \mathbf{H} = \frac{1}{2}
\mbox{tr}_{g}\mathbf{K}\in\mathfrak{X}^{\perp}(S). \]

{\noindent}The component of $\mathbf{H}$ along a given normal
direction $\eta$, denoted by $h_{\eta}$, is called the
\emph{expansion along $\eta$}, i.e., $h_{\eta} =
\widetilde{g}(\mathbf{H},\eta) = \mbox{tr}_{g}(A_{\eta})/2$.

Let us consider a local orthonormal basis $\{\eta_{1},\eta_{2}\}$
of the normal bundle of the spacelike surface $S$ in
$\mathbb{L}^{4}$, where $\eta_{1}$ is future-pointing timelike and
$\eta_{2}$ is spacelike. If we denote by $A_{i}$ the shape
operator associated with $\eta_{i}$, $i=1,2$, the shape tensor can
be written as

\[ \mathbf{K}(X,Y) = -g(A_{1}(X),Y)\eta_{1} +
g(A_{2}(X),Y)\eta_{2}, \]

{\noindent}for any tangent vector fields $X,Y$ to $S$. Assume that
$X(u,v)$ is a local parametrization on the surface $S$. Then, from
the classical theory of surfaces (see e.g. \cite{struik}), with
the notation $2 h_{i} = \mbox{tr}_{g}(A_{i})$ and

\begin{eqnarray*}
E = \widetilde{g}(X_{u},X_{u}), & F = \widetilde{g}(X_{u},X_{v}),
& G = \widetilde{g}(X_{v},X_{v}), \\
e_{i} = \widetilde{g}(X_{u u},\eta_{i}), & f_{i} =
\widetilde{g}(X_{u v},\eta_{i}), & g_{i} = \widetilde{g}(X_{v
v},\eta_{i}),
\end{eqnarray*}

{\noindent}we obtain

\[ 2 h_{i} = \frac{e_{i}G-2 f_{i}F + g_{i}E}{E G-F^{2}},\ \ \ i=1,2.
\]

{\noindent}Another useful local basis $\{\mathbf{k},\mathbf{l}\}$
of the normal bundle of $S$ can be chosen such that both vectors
are null, future-pointing and satisfy the normalization condition
$\widetilde{g}(\mathbf{k},\mathbf{l})=-1$. {\noindent}In the following, we choose

\begin{eqnarray*}
\mathbf{k} = \frac{1}{\sqrt{2}}(\eta_{1}-\eta_{2}) & \mbox{and} &
\mathbf{l} = \frac{1}{\sqrt{2}}(\eta_{1}+\eta_{2}).
\end{eqnarray*}

{\noindent}With respect to this normal basis the mean curvature
vector field $\mathbf{H}$ becomes

\[ \mathbf{H} = -\frac{\sqrt{2}}{2}(h_{1}+h_{2})\mathbf{k} -
\frac{\sqrt{2}}{2}(h_{1}-h_{2})\mathbf{l}. \]

{\noindent}In particular, the \emph{expansions along $\mathbf{k}$
and $\mathbf{l}$} are given by

\begin{eqnarray*}
\Theta_{k} = \frac{\sqrt{2}}{2}(h_{1}-h_{2}) & \mbox{and} &
\Theta_{l} = \frac{\sqrt{2}}{2}(h_{1}+h_{2}).
\end{eqnarray*}

Besides the extrinsic mean curvature, also the intrinsic Gaussian
curvature $K$ of the surface can be expressed in terms of the
coefficients of the first and second fundamental forms as (see
e.g. \cite{struik}),

\[ K = \frac{-\mbox{det}(A_{1}) +
\mbox{det}(A_{2})}{\mbox{det}(g)} = \frac{-e_{1}g_{1} + e_{2}g_{2}
+ f_{1}^{2} -f_{2}^{2}}{E G-F^{2}}. \]

A spacelike surface $S$ in $\mathbb{L}^{4}$ is called
\emph{invariant under spacelike rotations} if it is invariant
under the following group $\mathbf{G}_{s}$ of linear isometries of
$\mathbb{L}^{4}$:

\[ \mathbf{G}_{s} = \left\{ B_{\theta} = \left(
\begin{array}{cccc}
1 & 0 & 0 & 0 \\
0 & 1 & 0 & 0 \\
0 & 0 & \cos(\theta) & \sin(\theta) \\
0 & 0 & -\sin(\theta) & \cos(\theta) \\
\end{array} \right)\ :\ \theta\in\mathbb{R} \right\}, \]

{\noindent}i.e., if $B_{\theta}S = S$, for any
$\theta\in\mathbb{R}$.

Since we regard these surfaces as geometric objects, the main tool
to study them consist of introducing natural (local)
parameterizations, which can be achieved by making use of the
action of the group and finding a suitable profile curve. It is
worth pointing out that when we let a surface  be \textit{only} of
class $C^{\infty}$ and not analytical, we might get a very
complicated curve. More problems arise when the surface is
immersed, but not imbedded. Even worse, since the codimension is
two, the surface does not need to be orientable. As a consequence,
we will restrict our study to a local setting.

Note that the set of fixed points of
$\mathbf{G}_{s}$ is $\{(x_{1},x_{2},x_{3},x_{4})\in\mathbb{L}^{4}:
x_{3}=x_{4}=0\}$, so we need the following subset
$$\mathcal{P} =
\{(x_{1},x_{2},x_{3},x_{4})\in\mathbb{L}^{4}: x_{4}=0, x_{3}>0 \}.
$$
With the help of $\mathcal{P}$, we can introduce a parametrization $X(t,\theta)$
of $S$ as follows. Given a smooth curve
$\alpha: I\subset\mathbb{R}  \rightarrow \mathcal{P}$, $t  \mapsto  \alpha(t) = (\alpha_{1}(t),\alpha_{2}(t),\alpha_{3}(t),0)$, the parametrization can be written as
\[ X(t,\theta) = \Big(\alpha_{1}(t),\alpha_{2}(t),
\alpha_{3}(t)\cos(\theta),\alpha_{3}(t)\sin(\theta)\Big),\ \ \
t\in I, \theta\in\mathbb{R}. \] \noindent We denote by
$\Sigma_{\alpha}$ the parameterized surface associated with
$\alpha$, as a subset of $\Lfour$,
\[ \Sigma_{\alpha} = \{ X(t,\theta) = \alpha(t)\cdot B_{\theta}:
t\in I, \theta\in\mathbb{R}\}\subset S. \]
\noindent We recall that $\Sigma_{\alpha}$ might not cover the whole original surface $S$, but it would be a big enough open subset.
Next, without loss of generality we can assume that the
spacelike curve $\alpha$ is arc-length parameterized, i.e.,
$\widetilde{g}(\alpha'(t),\alpha'(t)) = 1$. The derivatives of
$X(t,\theta)$ are

\begin{eqnarray*}
X_{t} =
\Big(\alpha_{1}'(t),\alpha_{2}'(t),\alpha_{3}'(t)\cos(\theta),\alpha_{3}'(t)\sin(\theta)\Big)
& \mbox{and} & X_{\theta} =
\Big(0,0,-\alpha_{3}(t)\sin(\theta),\alpha_{3}(t)\cos(\theta)\Big).
\end{eqnarray*}

{\noindent}The Riemannian metric of the surface reads

\[ g = \mbox{d}t^{2} + \alpha_{3}^{2}\mbox{d}\theta^{2}. \]

{\noindent}A globally defined orthonormal tangent frame on
$\Sigma_{\alpha}$ is $u_{1} = X_{t}$ and $u_{2} =
X_{\theta}/\alpha_{3}$, and a globally defined orthonormal
basis of the normal bundle of $\Sigma_{\alpha}$ is given by

\begin{eqnarray*}
\eta_{1} & = & \frac{1}{\sqrt{1+(\alpha_{1}')^{2}}}\Big(
1+(\alpha_{1}')^{2},
\alpha_{1}'\alpha_{2}',\alpha_{1}'\alpha_{3}'\cos(\theta),
\alpha_{1}'\alpha_{3}'\sin(\theta)\Big), \\
\eta_{2} & = & \frac{1}{\sqrt{1+(\alpha_{1}')^{2}}}\Big( 0,
-\alpha_{3}',
\alpha_{2}'\cos(\theta),\alpha_{2}'\sin(\theta)\Big),
\end{eqnarray*}

{\noindent}with $\eta_{1}$ future-pointing timelike and $\eta_{2}$
spacelike. A straightforward computation shows that the components
of the second fundamental form are given by

\begin{eqnarray} \label{eq:efg}
e_{1} = -\frac{\alpha_{1}''}{\sqrt{1+(\alpha_{1}')^{2}}}, & f_{1}
= 0, & g_{1} = -\frac{\alpha_{1}'\, \alpha_{3}\,
\alpha_{3}'}{\sqrt{1+(\alpha_{1}')^{2}}},
\\ \nonumber
e_{2} = \frac{\alpha_{2}'\, \alpha_{3}'' -\alpha_{2}''\,
\alpha_{3}'}{\sqrt{1+(\alpha_{1}')^{2}}}, & f_{2} = 0, & g_{2} =
-\frac{\alpha_{2}'\, \alpha_{3}}{\sqrt{1+(\alpha_{1}')^{2}}}.
\end{eqnarray}

{\noindent}Hence the shape operators associated with $\eta_{1}$
and $\eta_{2}$ are simultaneously diagonalizable, i.e. the normal
curvature $R^{\perp}$ of the normal bundle vanishes identically.
The expansions along $\eta_{1}$ and $\eta_{2}$ are

\begin{eqnarray}
2 h_{1} = -\frac{\alpha_{1}'\, \alpha_{3}' + \alpha_{3}\,
\alpha_{1}''}{\alpha_{3}\sqrt{1+(\alpha_{1}')^{2}}} & \mbox{and} &
2 h_{2} = - \frac{\alpha_{2}'+\alpha_{3}(\alpha_{2}''\,
\alpha_{3}'-\alpha_{2}'\,
\alpha_{3}'')}{\alpha_{3}\sqrt{1+(\alpha_{1}')^{2}}}.
\label{eq:meancurv1}
\end{eqnarray}

{\noindent}The Gaussian curvature of a spacelike surface which is
invariant under a spacelike rotation is

\begin{equation}
K = -\frac{\alpha_{3}''}{\alpha_{3}}. \label{eq:gauss1}
\end{equation}


\section{Classification theorem and a gluing procedure}

The following classification is local, i.~e., a surface $S$ which
is invariant by $\mathbf{G}_s$ will be locally congruent to the
surfaces in the next theorem.

\begin{theorem} \label{th:class1} Let $\Sigma_{\alpha}$ be a surface in $\Lfour$ which is invariant under
spacelike rotations. Assume that its mean curvature vector
satisfies $\|\mathbf{H}\|=0$. Then, the generating curve
$\alpha(t)=(\alpha_1(t),\alpha_2(t),\alpha_3(t),0)$ is locally described by one
of the following cases:
\begin{description}
\item[\textit{A})] Given a smooth function $\tau:I\subset (0,\infty)\longrightarrow\mathbb{R}$,
choose  a function $\varepsilon:I\longrightarrow\{1,-1\}$ such
that $\varepsilon \tau$ is also smooth. Define the coordinate
functions $\alpha_i:I\longrightarrow\mathbb{R}$, $i=1,2,3$, as
follows

\begin{equation}
\alpha_1(t)=\int \varepsilon(t)\tau(t){\rm d}t, \quad \alpha_2(t)
=\int \tau(t){\rm d}t, \quad \alpha_3(t)=t. \label{eq:theorcaseA}
\end{equation}
Moreover, the mean curvature vector of $\Sigma_{\alpha}$ is
$$ \mathbf{H}=\frac{\tau+t\tau'}{2t\sqrt{1+\tau^2}}\left( \varepsilon \eta_1-\eta_2 \right).
$$

\item[\textit{B})] Given a smooth positive function $\alpha_{3}:
I\subset\mathbb{R}\rightarrow{\mathbb{R}}$, and two constants
$\varepsilon_{1},\varepsilon_{2}=\pm 1$, define the functions
\begin{equation}
\xi(t)=\int\frac{\mathrm{d}t}{\alpha_{3}(t)}, \quad
\alpha_{1}(t) =  \varepsilon_{1}\,\int \left\{
\sinh(\xi(t)) - \alpha_{3}'(t)\,\cosh(\xi(t))\right\}\,
\mathrm{d}t, \label{eq:theorem1}
\end{equation}
{\noindent}and
\begin{equation}
\alpha_{2}(t) = \varepsilon_{2}\,\int\left\{
\cosh(\xi(t)) - \alpha_{3}'(t)\,
\sinh(\xi(t))\right\}\,
\mathrm{d}t. \label{eq:theorem2}
\end{equation}
Moreover, the mean curvature vector of $\Sigma_{\alpha}$ is
$$ \mathbf{H} = \frac{\cosh(\xi(t))\left( 1-\alpha_{3}'(t)^{2} -\alpha_{3}(t)\,
\alpha_{3}''(t)\right)}{2\alpha_{3}(t)\sqrt{1+\alpha_{1}'(t)^{2}}}
\left(\varepsilon_1\eta_1-\varepsilon_2\eta_2 \right).
$$
\end{description}
\noindent In addition, in Case B, given two unit curves
$\alpha(t)=(\alpha_1(t),\alpha_2(t),\alpha_3(t),0)$ and $\beta(t)
= (\beta_{1}(t),\beta_{2}(t),$ $\beta_{3}(t),0)$, such that
$\alpha_{3}(t)=\beta_{3}(t)$, there exists an affine isometry $F$
of $\mathbb{L}^{4}$ satisfying
$F(\Sigma_{\alpha})=\Sigma_{\beta}$.
\end{theorem}

\begin{proof}
We recall the generating spacelike unit curve
$\alpha:J\longrightarrow \mathcal{P}$,
$\alpha(t)=(\alpha_1(t),\alpha_2(t),$ $\alpha_3(t),0)$ of
$\Sigma_{\alpha}$, with $\alpha_3(t)>0$. Now, we consider two
subsets $J_0=\{t\in J : \alpha_3'(t)=\pm 1\}$ and $J_1=\{t\in J :
\alpha_3'(t)\neq \pm 1\}$, which are not intervals in general.
Since $\alpha_3'$ is continuous, $J_1$ is an open subset of $J$,
i.~e., it is either the empty set or made of countable many open
intervals.  From a topological point of view, $J_1$ might be empty
or a mixture of intervals, accumulation points and isolated
points. To make some progress, we need to work on open intervals
included in $J_0$ and $J_1$.
\par
\noindent \textit{Case A.} We assume that there exists an open
interval $I\subset J_0$ where $\alpha_3'(t)^2=1$.
By a change of parameter, we can assume without loss of generality
that $I\subset (0,\infty)$ and $\alpha_3(t)=t$ on $I$. From now
on, we work on $I$. Since $\alpha$ is unit, we know that
$\alpha_1'(t)^2=\alpha_2'(t)^2$. Thus, there exists a function
$\varepsilon:I\longrightarrow\{-1,1\}$ such that $\varepsilon\,
\alpha_1'(t)=\alpha_2'(t)$. Now, equations (\ref{eq:theorcaseA})
are trivially satisfied.

\par
\noindent\textit{Case B.} We assume that there exists an open
interval $I\subset J_1$, so we work on $I$. Since $\alpha$ is
arc-length parameterized, we have
$(-\alpha_{1}'+\alpha_{2}')(\alpha_{1}'+\alpha_{2}')=1-(\alpha_{3}')^2$.
By shrinking $I$ if necessary, we can introduce an \emph{angle}
function $\xi(t)$ and a constant $\varepsilon=\pm 1$, such that
$$
 -\alpha_{1}'(t)+\alpha_{2}'(t)  = \varepsilon\, (1+\alpha_{3}'(t))\, \exp(\xi(t)), \quad
 \alpha_{1}'(t)+\alpha_{2}'(t)  = \varepsilon\, (1-\alpha_{3}'(t))\, \exp(-\xi(t)).
$$
In this way, we obtain the following expressions:
\begin{eqnarray}
\alpha_{1}'(t) &=& \frac{1}{2}\Big\{ \varepsilon\, (1-\alpha_{3}'(t))\, \exp(\xi(t)) - \varepsilon\, (1+\alpha_{3}'(t))\, \exp(-\xi(t))\Big\}, \label{eq:proofxi1} \\
\alpha_{2}'(t) &=& \frac{1}{2}\Big\{ \varepsilon\,
(1-\alpha_{3}'(t))\, \exp(\xi(t)) +\varepsilon\,
(1+\alpha_{3}'(t))\, \exp(-\xi(t))\Big\}. \label{eq:proofxi2}
\end{eqnarray}
Since we are assuming $\|\mathbf{H}\|=0$, there exists a function
$\delta:I\longrightarrow\{1,-1\}$ such that $h_{1}=\delta\,
h_{2}$. Bearing in mind (\ref{eq:proofxi1}) and
(\ref{eq:proofxi2}), we substitute this in (\ref{eq:meancurv1}),
obtaining
\begin{eqnarray}
& (1+ \delta(t))\, \alpha_{3}'(t)\,
\Big(1- \alpha_{3}(t)\, \xi'(t)\Big)\, \sinh(\xi(t)) & \label{eq:h1-delta-h2} \\
& +\Big\{ \alpha_{3}'(t)^2\,
\Big(\delta(t)\alpha_{3}(t)\xi'(t)-1\Big) +\alpha_{3}(t)\,
\Big(\xi'(t) +(\delta(t)-1)\, \alpha_3''(t)\Big)-\delta(t)
\Big\}\, \cosh(\xi(t))\, =\, 0. &\nonumber
\end{eqnarray}
Now, two cases arise naturally.
\begin{enumerate}
\item  We suppose that there exists an open interval $I^{+}$ such that $\delta(t)=1$ for any $t\in I^{+}$. We work in this interval. Equation (\ref{eq:h1-delta-h2}) becomes
$$ 0\ =  \Big(\cosh(\xi(t))\, \alpha_3'(t)^2-2\sinh(\xi(t))\, \alpha_3'(t)+\cosh(\xi(t))\Big)\, \Big(\alpha_3(t)\xi'(t)-1\Big).
$$
Now, we suppose that there exists a $t_0\in I^{+}$ such that
$0=\cosh(\xi(t_0))\, \alpha_3'(t_0)^2-2\sinh(\xi(t_0))\,
\alpha_3'(t_0)+\cosh(\xi(t_0))$. However, from this
equation, we obtain $\alpha_3'(t_0)= \tanh(\xi(t_0))$
$\pm \sqrt{-1}\,{\rm sech}(\xi(t_0))$, which is impossible. Thus,
on the whole $I^{+}$ (at least), we obtain
\[ \xi(t) = \int\frac{\mathrm{d}t}{\alpha_{3}(t)}. \]
\noindent Inserting this in (\ref{eq:proofxi1}) and
(\ref{eq:proofxi2}) gives the expressions (\ref{eq:theorem1}) and
(\ref{eq:theorem2}) for the case  $\varepsilon_1=\varepsilon_2=\varepsilon$.

\item We suppose that there exists an open interval $I^{-}$ such that $\delta(t)=-1$ for any $t\in I^{-}$. We work in this interval. Equation (\ref{eq:h1-delta-h2}) becomes
\[-\alpha_3'(t)^2\, \Big(\alpha_3(t)\xi'(t)+1\Big)+\alpha_3(t)\, \Big(\xi'(t)-2\alpha_3''(t)\Big)+1=0.\]
From here, we compute $\xi'(t)= \frac{-1}{\alpha_3(t)} - \,\frac{2\alpha_3''(t)}{\alpha_3'(t)^2-1}.$ Now, we obtain
$$ \xi(t) =  -\int\frac{\mathrm{d}t}{\alpha_{3}(t)} -
\ln\left|\frac{\alpha_{3}'(t)-1}{\alpha_{3}'(t)+1}\right|.
$$

{\noindent}When inserting this expression in (\ref{eq:proofxi1}) and
(\ref{eq:proofxi2}), one cannot forget the signs, i.~e., $(1+\alpha_3')\exp\left(\ln\Big\vert\frac{\alpha_3'-1}{1+\alpha_3'}\Big\vert\right) = \pm (\alpha_3'-1)$. Bearing this in mind,  two cases arise. However, it is possible to deal with both at the same time by choosing suitable constants $\varepsilon_1, \varepsilon_2=\pm 1$,  obtaining again expressions (\ref{eq:theorem1}) and (\ref{eq:theorem2}). Thus, there is no loss of generality if we redefine the angle function as $\xi(t)=\int (1/\alpha_3(t)){\rm d}t$.

\end{enumerate}

\noindent Let $\beta(t)$ be another arc-length parameterized spacelike
curve, with $\beta_{3}(t)=\alpha_{3}(t)$. Then, we have

\[ \int\frac{\mathrm{d}t}{\beta_{3}(t)} =
\int\frac{\mathrm{d}t}{\alpha_{3}(t)} + c_{0}, \]

{\noindent}with $c_{0}$ an integration constant. A straightforward
computation shows

\[ (\beta_{1}',\beta_{2}') =
(\alpha_{1}',\alpha_{2}')\left(\begin{array}{cc}
                               \widetilde{\varepsilon}_{1} & 0 \\
                               0 & \widetilde{\varepsilon}_{2} \\
                               \end{array}\right) \left(
                               \begin{array}{cc}
                               \cosh(c_{0}) & \sinh(c_{0}) \\
                               \sinh(c_{0}) & \cosh(c_{0}) \\
                               \end{array}\right), \]

{\noindent}with $\widetilde{\varepsilon}_{1},
\widetilde{\varepsilon}_{2}= \pm 1$. If the integration constants
of (\ref{eq:theorem1}) and (\ref{eq:theorem2}) are denoted by
$\alpha_{1}^{0}$ and $\alpha_{2}^{0}$, we call
$v=(\alpha_{1}^{0},\alpha_{2}^{0},0,0)$. The affine isometry
$F:\mathbb{L}^{4}\rightarrow \mathbb{L}^{4}$,

\[ F(x_{1},x_{2},x_{3},x_{4}) =
(x_{1},x_{2},x_{3},x_{4})\left(\begin{array}{cccc}
                               \widetilde{\varepsilon}_{1} & 0 & 0 & 0 \\
                               0 & \widetilde{\varepsilon}_{2} & 0 & 0
                               \\
                               0 & 0 & 1 & 0 \\
                               0 & 0 & 0 & 1 \\
                               \end{array}\right) \left(
                               \begin{array}{cccc}
                               \cosh(c_{0}) & \sinh(c_{0}) & 0
                               & 0 \\
                               \sinh(c_{0}) & \cosh(c_{0}) & 0
                               & 0 \\
                               0 & 0 & 1 & 0 \\
                               0 & 0 & 0 & 1 \\
                               \end{array} \right) + v, \]

{\noindent}satisfies $F\circ \alpha = \beta$ and thus
$F(\Sigma_{\alpha}) = \Sigma_{\beta}$. \qed
\end{proof}

\vspace{4mm}

In Cases A and B of the previous theorem, the domain of the curve
$\alpha$ might not be connected. If we ask the domain $I$ of
$\alpha$ to be an interval, we will say that the surface
$\Sigma_{\alpha}$  is of \textit{type A} or \textit{type B},
according to Cases A or B, respectively. In particular, surfaces
of type A and B have to be connected, orientable, and any normal
lightlike vector can be globally defined.

\begin{corollary} \begin{enumerate}
\item A surface of type A is a MOTS if, and only if, the function $\varepsilon$ is a global constant. In addition, a surface of type $A$ is marginally trapped if, and only if, the function $\varepsilon$ is a global constant and $\tau(t)+t\tau'(t)$ is globally positive or negative.
\item Any surface of type B is a MOTS. In addition, a surface of type B is marginally trapped if, and only if, the function $1-(\alpha_3')^2-\alpha_3\alpha_3''$ is globally positive or negative.
\end{enumerate}
\end{corollary}

By a result in \cite{penrose} (see also \cite{mars}), a closed surface of type A or B bounding a domain cannot exist.  Thus, a good second alternative is completeness.

\begin{corollary} Let $\Sigma_{\alpha}$ be a surface of type B in $\mathbb{L}^{4}$. If
$\alpha_3:\mathbb{R}\longrightarrow\mathbb{R}$ is a smooth
function such that $\alpha_3(t)\geq a_0>0$ for some real constant
$a_0$, then the surface $\Sigma_{\alpha}$ is complete.
\end{corollary}

\begin{proof} The metric of the surface $\Sigma_{\alpha}$ of type B
satisfies
$$ g \geq {\rm d}t^2+a_0^2{\rm d}\theta^2,$$
and it is defined for any $t,\theta\in\mathbb{R}$. This means that
$\Sigma_{\alpha}$ is complete. \qed
\end{proof}

From the proof of Theorem \ref{th:class1}, we find a characterization of the extremal spacelike surfaces which are invariant under a spacelike rotation.

\begin{corollary}\label{cor:extremal}
A spacelike surface in $\Lfour$ is extremal and invariant under a spacelike rotation if, and only if, it is locally congruent to a surface $\Sigma_{\alpha}$ whose profile curve
$\alpha:I\subset (0,\infty)\rightarrow\mathcal{P}$, $\alpha(t)=(\alpha_1(t),\alpha_2(t),\alpha_3(t),$ $0)$, is given by one of the following cases:
\begin{enumerate}
\item \label{AyB}$\alpha_1(t) = a\, \varepsilon_1 \ln (t) +b$, $\alpha_2(t)= a\, \varepsilon_2\ln (t) +c$, $\alpha_3(t)=t$, with $a>0$, $b,c\in\mathbb{R}$ and $\varepsilon_1,\varepsilon_2=\pm 1$.
\item $\alpha_{1}(t)  =
\frac{\varepsilon_{1}}{2}(a^{2}+1-b)\,\ln\left| a+t+\sqrt{t^{2}+2at+b}\right|$, \\ 
$\alpha_{2}(t)  =   \frac{\varepsilon_{2}}{2}(a^{2}-1-b)\,\ln\left|a+t+\sqrt{t^{2}+2at+b}\right|$,\\
$\alpha_{3}(t)  =  \sqrt{t^{2}+2a t+b},$
{\noindent} where
$\varepsilon_{1},\varepsilon_{2}=\pm 1$, and
$a,b\in\mathbb{R}$.
\end{enumerate}
\end{corollary}

\begin{remark}\label{rem:caseAB} Surfaces of type A and B are not excluding. Indeed, by considering $\alpha_3(t)=t$ in Theorem \ref{th:class1}, we obtain Case \ref{AyB} of Corollary \ref{cor:extremal}, which is a description of all curves generating surfaces which are simultaneously of type A and B.
\end{remark}

\begin{remark} All surfaces of type A are flat, i.e. their Gaussian curvature $K=0$.
\end{remark}

\begin{remark} Given a surface of type A, if the mean curvature vector is future-pointing, by considering the function  $-\varepsilon$, we obtain a surface with past-pointing mean curvature vector, and viceversa. A similar situation holds for surfaces of type B by changing $\varepsilon_1$ by $-\varepsilon_1$.
\end{remark}

\begin{remark}\label{rem:isometry} Given a surface of type A, with the very same function $\tau$ (and same function $\varepsilon$)  it is possible to construct infinitely many curves, and thus surfaces of type A. However, two of them are related by the translation defined by considering different integration constants in the expressions of functions $\alpha_1$ and $\alpha_2$. In addition, if we change $\varepsilon$ by $-\varepsilon$, the reflection by a suitable hyperplane links both surfaces.
\end{remark}

\begin{remark} For a surface of type A, with the function $\tau(t)=
c\in \mathbb{R}$, both shape operators $A_{1}$ and $A_{2}$ are of
rank 1. Such a surface is called pseudo-isotropic. See e.g.
\cite{rosca1} for properties of such surfaces.
\end{remark}

\begin{remark} \textit{Surfaces of type A as graphs, locally.}
Given a surface $S$ of type A, with the function $\varepsilon$
locally constant. We restrict this remark to an interval $J$ where
$\varepsilon$ is constant. Then,  the surface is included in the
null hyperplane $\mathcal{H}=\{(x_1,x_2,x_3,x_4)\in\Lfour :
x_1=\varepsilon x_2\}$. From a Set Theory point of  view, one can
identify $\mathcal{H}$ with $\mathbb{R}^3$, where the surface is
parameterized as $Y(t,\theta)=(\int \tau(t){\rm d}t,t \cos\theta,
t\sin\theta)$. We just call $T(t)=\int \tau(t){\rm d}t$, so we can
identify a region of $S$ with  the set $\{(T(\sqrt{y^2+z^2}),y,z)
: \sqrt{y^2+z^2}\in J\}$. Conversely, any surface of type A can be
locally seen as a graph over an annulus centered at the origin of
$\mathbb{R}^2$. Furthermore, given a real constant $a>0$,  a disk
$D(a)=\{(y,z)\in\mathbb{R}^2: y^2+z^2<a^2\}$ and a smooth function
$T:D(a)\longrightarrow\mathbb{R}$, such that $T$ is invariant by
transformations of the form $(y,z)\mapsto
(y\cos\theta-z\sin\theta,y\sin\theta+z\cos\theta)$,
$\theta\in\mathbb{R}$, then the graph $S=\{(T(y,z),y,z): (y,z)\in
D(a)\}$ can be imbedded in $\Lfour$ as a surface whose  mean
curvature vector field satisfies $\|\mathbf{H}\|=0$, and admitting
a parametrization of a surface of type $A$ except in the point
touching the plane of fixed points of $\mathbf{G}_s$.
\end{remark}

We consider two bounded spacelike surfaces of type A and B, and
suppose that the mean curvature vector of a surface of type A is
constantly either past or future-pointing near one of its
boundaries. In such case, we describe a method to glue them in one
new spacelike surface which is invariant by a spacelike rotation
with an intermediate region satisfying $\mathbf{H}=0$.

\begin{proposition}\label{pr:glue1} Let $\Sigma_{\alpha}$ and $\Sigma_{\beta}$ be two surfaces of type A and B as
in Theorem \ref{th:class1} with generating curves
$\alpha:(a,b)\longrightarrow\mathcal{P}$,
$\beta:(c,d)\longrightarrow\mathcal{P}$, with $0\leq a<b<c<d\leq
\infty$. Assume that there is a constant $\omega>0$ such that the
function $\varepsilon$ is constant on the interval $(b-\omega,b)$.
Then, there exists an affine isometry
$F:\Lfour\longrightarrow\Lfour$, a real number $\nu>0$ and a
unit spacelike curve $\gamma:(a,d)\longrightarrow\mathcal{P}$,
satisfying that the surface $\Sigma_{\gamma}$ is invariant by a
spacelike rotation,
$\gamma\vert_{(a,b-\nu)}=\alpha\vert_{(a,b-\nu)}$,
$F(\Sigma_{\gamma\vert_{(c+\nu,d)}})=\Sigma_{\beta\vert_{(c+\nu,d)}}$ and the mean curvature of the region $\Sigma_{\gamma\vert_{(b,c)}}$ vanishes identically.
\end{proposition}

\begin{proof} Because the function $\varepsilon$ is constant on the interval $(b-\omega,b)$,
there is no loss of generality if we assume
$\varepsilon(t)=\varepsilon_1\varepsilon_2$ for any $b-\omega<t<b$
(by changing $\varepsilon_1$). In such case, the surface
$\Sigma_{\beta}$ is unique up to an affine isometry as in Theorem
\ref{th:class1}.

We choose $\nu\in\mathbb{R}$ such that
$0<\nu<\min\{(d-c)/4,(b-a)/4,\omega\}$ and consider two smooth
functions $f_i:(a,d)\longrightarrow\mathbb{R}$, $i=1,2$,
satisfying
\begin{enumerate}
 \item $0\leq f_i\leq 1$, $i=1,2$;
\item $f_1(t)=0$ and $f_2(t)=1$ for any $t\in(a,c)$;
\item $f_1(t)=1$ and $f_2(t)=0$ for any $t\in(c+\nu,d)$.
\end{enumerate}
Note that $f_1'=f_2'=0$ on the intervals $(a,c)$ and $(c+\nu,d)$.
We define the smooth function
$\gamma_3:(a,d)\longrightarrow(0,\infty)$, given by
$\gamma_3(t)=tf_2(t)+\beta_3(t)f_1(t)$. It is straightforward to
check that $\gamma_3(t)=t$ for any $t\in(a,c)$ and
$\gamma_3(t)=\beta_3(t)$ for any $t\in(c+\nu,d)$. In particular,
$\gamma_3(t)=\alpha_3(t)$ on the interval $(a,b)$.

We define $\tilde{\xi}(t)=\int \frac{{\rm d}t}{\gamma_3(t)}$, with
the additional condition $\tilde{\xi}(t)=\xi(t)$ for any
$t\in(c+\nu,d)$, which can be achieved by choosing a suitable
integration constant. Bearing in mind Case B in Theorem
\ref{th:class1}, we define $\tilde{\beta}_1,
\tilde{\beta}_2:(a,d)\longrightarrow\mathcal{P}$, satisfying
$\tilde{\beta}_1(t)=\beta_1(t)$ and
$\tilde{\beta}_2(t)=\beta_2(t)$ for any $t\in(c+\nu,d)$.

Next, we consider two smooth functions $f_3,f_4:(a,d)\longrightarrow\mathbb{R}$ such that
\begin{enumerate}
 \item $0\leq f_i\leq 1$, $i=3,4$;
\item $f_3(t)=1$ and $f_4(t)=0$ for any $t\in(a,b-\nu)$;
\item $f_3(t)=0$ and $f_4(t)=1$ for any $t\in(b,d)$.
\end{enumerate}
Let $\tau(t)$ be the function in the definition of the curve
$\alpha$. Next, we define the smooth functions
$\gamma_i:(a,d)\longrightarrow\mathbb{R}$, $i=1,2$, given by
$$\gamma_1(t) =\int\Big(\varepsilon(t)\, \tau(t)\, f_3(t)+\tilde{\beta}'_1(t)\, f_4(t)\Big)\, {\rm d}t, \quad
\gamma_2(t) =\int\Big( \tau(t)\, f_3(t)+\tilde{\beta}'_2(t)\,
f_4(t)\Big)\, {\rm d}t,$$ but satisfying $\gamma_1(t)=\alpha_1(t)$
and $\gamma_2(t)=\alpha_2(t)$ for any $t\in (a,b-\nu)$. As above,
it is only necessary to choose suitable integration constants.
Indeed, given $t\in(a,b-\nu)$, $\gamma_1'(t)=\varepsilon(t)\,
f_3(t)\, \tau(t)+f_4(t)\, \tilde\beta'_1(t)=\varepsilon(t)\,
\tau(t)$, and $\gamma_2'(t)=f_3(t)\, \tau(t)+f_4(t)\,
\tilde\beta'_2(t)=\tau(t)$. Next, given $t\in(b,d)$,
$\gamma_i'(t)=\tilde\beta_i'(t)$, for $i=1,2$, and
$\gamma_3(t)=t$. Note that by Corollary \ref{cor:extremal}, the surface
$\Sigma_{\gamma\vert_{(b,c)}}$ satisfies $\mathbf{H}=0$. Now,
bearing in mind Remark \ref{rem:caseAB}, given $t\in(b-\nu,b)$, we
see $\tilde\beta_i'(t)=2\varepsilon_i\, \exp(\xi_0)\, t$, $i=1,2$.
Thus, since $0<\nu<\omega$, $\gamma_1'(t)=\varepsilon(t)\,
\Big(f_3(t)\, \tau(t)+f_4(t)\,
\frac{\varepsilon_1}{\varepsilon(t)}\exp(\xi_0)\, t\Big)=
\varepsilon(t)\, \Big(f_3(t)\, \tau(t)+f_4(t)\, \varepsilon_2\,
\exp(\xi_0)\, t\Big)=\varepsilon(t)\, \gamma_2'(t)$. Finally, we
define the curve
$$ \gamma:(a,d)\longrightarrow\mathcal{P}, \quad \gamma(t)=(\gamma_1(t),\gamma_2(t),\gamma_3(t),0),$$
and its associated surface $\Sigma_{\gamma}$. From the above
computations, it is easy to check that $\|\gamma'\|=1$. Needless
to say, $\Sigma_{\alpha}$ is an open subset of $\Sigma_{\gamma}$.
It only remains to point out that the open subset
$\Sigma_{\gamma\vert_{(c+\nu,d)}}$ of $\Sigma_{\gamma}$ might not
be the original $\Sigma_{\beta\vert_{(c+\nu,d)}}$, but they will
be congruent by an affine isometry, as in Theorem \ref{th:class1}.
\qed
\end{proof}

\begin{corollary} Let $\Sigma_{\alpha}$ and $\Sigma_{\beta}$ two surfaces of type A,
whose generating curves
$\tau_{\alpha}:(a,b)\longrightarrow\mathbb{R}$ and
$\tau_{\beta}:(c,d)\longrightarrow\mathbb{R}$ satisfy
$0<a<b<c<d\leq \infty$. Then, there exists a unit spacelike curve
$\gamma:(a,d)\longrightarrow\mathcal{P}$, a real number $\nu>0$
and two translations $F_{\alpha},
F_{\beta}:\Lfour\longrightarrow\Lfour$  such that
$\Sigma_{\gamma}$ is a surface of type A,
$F_{\alpha}(\Sigma_{\gamma\vert_{(a,b-\nu)}})=\Sigma_{\alpha\vert_{(a,b-\nu)}}$,
$F_{\beta}(\Sigma_{\gamma\vert_{(c+\nu,d)}})=\Sigma_{\beta\vert_{(c+\nu,d)}}$ and the mean curvature of the region $\Sigma_{\gamma\vert_{(b,c)}}$ vanishes identically.
\end{corollary}

\begin{corollary} Let $\Sigma_{\alpha}$ and $\Sigma_{\beta}$ two surfaces of type B, with profile curves   $\alpha:(a,b)\longrightarrow\mathcal{P}$ and $\beta:(c,d)\longrightarrow\mathcal{P}$, $-\infty\leq a<b<c<d\leq \infty$. Then, there exists a unit spacelike curve $\gamma:(a,d)\longrightarrow\mathcal{P}$, a real number $\nu>0$ and two affine isometries $F_{\alpha}, F_{\beta}:\Lfour\longrightarrow\Lfour$ such that $\Sigma_{\gamma}$ is a surface of type B, $F_{\alpha}(\Sigma_{\gamma\vert_{(a,b-\nu)}})=\Sigma_{\alpha\vert_{(a,b-\nu)}}$,  $F_{\beta}(\Sigma_{\gamma\vert_{(c+\nu,d)}})=\Sigma_{\beta\vert_{(c+\nu,d)}}$ and the mean curvature of the region $\Sigma_{\gamma\vert_{(b,c)}}$ vanishes identically..
\end{corollary}
\noindent All necessary ideas to prove these two corollaries are
contained in the proof of Proposition \ref{pr:glue1} and in Remark
\ref{rem:isometry}.

\begin{remark} The methods explained in Proposition \ref{pr:glue1} and its two corollaries
give the possibility to construct surfaces $S$ satisfying the
following conditions:
\begin{enumerate}
\item $S$ is invariant under a spacelike rotation group.
 \item The mean curvature vector of $S$ satisfies $\|\mathbf{H}\|=0$, with (infinitely many countable) regions $\{S_n: n\in N\subset\mathbb{N}\}$ where its mean curvature vector $\mathbf{H}\neq 0$.
\item Each region $S_n$ can be either of type A or B.

\item The mean curvature vector of each region $S_n$ can be set either future or past-pointing, as \textit{desired}.
\item Among two \textit{adjacent} regions $S_n$ and $S_{n+1}$, there is an open
subset which is extremal, i.e. $\mathbf{H}=0$.
\end{enumerate}
In particular, it is possible to construct examples of several of
the types given in the classification introduced in \cite{senovilla}.
\end{remark}


\section{The Gaussian Curvature}

We show that there exist surfaces in $\Lfour$ invariant by
$\mathbf{G}_s$, whose mean curvature vector field satisfies $\|
\mathbf{H}\|=0$ and  with prescribed Gaussian curvature. As an
application, we compute all such surfaces which have constant
Gaussian curvature.

\begin{corollary}
Let $\kappa: I\subset\mathbb{R}\rightarrow\mathbb{R}$ be a smooth
function and $t_{0}\in I$. There exist $\delta>0$ and a unit curve
$\alpha: (t_{0}-\delta,t_{0}+\delta)\subset\mathbb{R}\rightarrow\mathcal{P}$, such that
$\alpha$ is a profile curve of a spacelike surface $\Sigma_{\alpha}(t,\theta)$ of type B  and whose Gaussian curvature at every
point $(t,\theta)$ is $\kappa(t)$. Moreover, if $\kappa(t)\,
\alpha_{3}(t)^{2} -\alpha_{3}'(t)^{2}+1$ never vanishes on $(t_{0}-\delta,t_{0}+\delta)$,  the surface $\Sigma_{\alpha}$ is marginally trapped.
\end{corollary}

\begin{proof} Given a smooth function $\kappa:
I\subset\mathbb{R}\rightarrow\mathbb{R}$ and let $\alpha_{3}:
(t_{0}-\delta,t_{0}+\delta)\rightarrow\mathcal{P}$ be a positive
solution of the differential equation $\alpha_{3}''(t) =
-\kappa(t)\, \alpha_{3}(t)$ (see (\ref{eq:gauss1})). The result
then follows from Theorem~\ref{th:class1}. \qed
\end{proof}

\begin{corollary}
There do not exist extremal spacelike surfaces of type B with constant
Gaussian curvature in $\mathbb{L}^{4}$.
\end{corollary}

\begin{proof} If we take $\kappa$ constant in the previous Corollary, the
solution of the differential equation $\kappa\, \alpha_{3}(t)^{2}
+\alpha_{3}'(t)^{2}-1= 0$ is either $\alpha_{3}(t) =
\varepsilon/\sqrt{-\kappa}$ if $\kappa<0$ or $\alpha_{3}(t) =
\varepsilon \sinh( (t-c)\sqrt{\kappa})/\sqrt{\kappa}$ if
$\kappa>0$, with $\varepsilon=\pm 1$ and $c\in\mathbb{R}$.
By (\ref{eq:gauss1}), using
these expressions in the differential equation $\alpha_{3}''(t) =
-\kappa\, \alpha_{3}(t)$ gives a contradiction in both cases. \qed
\end{proof}

\begin{example}\label{ej:1}
A surface of type B is flat if and only if a profile curve $\alpha:
(-b/a,+\infty)\subset\mathbb{R}\rightarrow\mathcal{P}$, with $a,
b\in\mathbb{R}$, $\mid\!a\!\mid\neq 0,1$, is given by

\begin{eqnarray*}
\alpha_{1}(t) & = & \frac{\varepsilon}{2}\left\{ \frac{1-a}{1+a}
(a t+b)^{\frac{a+1}{a}} + \frac{1+a}{1-a} (a
t+b)^{\frac{a-1}{a}}\right\}, \\
\alpha_{2}(t) & = & \frac{\varepsilon}{2}\left\{ \frac{1-a}{1+a}
(a t+b)^{\frac{a+1}{a}} - \frac{1+a}{1-a} (a
t+b)^{\frac{a-1}{a}}\right\}, \\
\alpha_{3}(t) & = & a t+b,
\end{eqnarray*}

{\noindent}or a profile curve $\alpha:
I\subset\mathbb{R}\rightarrow\mathcal{P}$ is given by

\[ \alpha(t) = \left( \varepsilon_{1}\, b\,
\cosh\left(\frac{t}{b}\right), \varepsilon_{2}\, b\,
\sinh\left(\frac{t}{b}\right), b,0\right), \]

{\noindent}with $b\in\mathbb{R}_{0}^{+}$.
\end{example}

\begin{example} \label{ej:2} Given $K>0$, we compute the profile curve $\alpha$ of a surface of type B with constant
Gaussian curvature $K^2$. By (\ref{eq:gauss1}), we need to solve
the differential equation $\alpha_3''(t)=-K^2\alpha_3(t)$, whose
general solution is
$$ \alpha_3(t) = c_1 \cos(K t+c_2), \quad \textrm{with} \ c_1,c_2\in\mathbb{R}, \ c_1\neq 0. $$
As $\alpha_3(t)$ has to be positive, we can choose
$I=(-\frac{\pi+2c_2}{2K},\frac{\pi-2c_2}{2K})$ if $c_1>0$ or
$I=(\frac{\pi-2c_2}{2K},\frac{3\pi-2c_2}{2K})$ if $c_1<0$, as the
domain of $\alpha_3(t)$. According to Theorem \ref{th:class1}, we
need to compute a primitive of $1/\alpha_3(t)$, which is
$$\xi(t)=\frac{1}{c_1} \ln\left\vert \frac{1+\sin(K t+c_2)}{1-\sin(K t+c_2)}\right\vert +\xi_0,$$
being $\xi_0\in\mathbb{R}$. This way, by taking $\varepsilon_1$,
$\varepsilon_2=\pm 1$, the coordinate functions of
$\alpha(t)=(\alpha_1(t),\alpha_2(t),\alpha_3(t),0)$ are
\begin{eqnarray*}
\alpha_1(t)&=&\varepsilon_1\Big( \sinh(\xi(t))+c_1\sin(K t+c_2)\cosh(\xi(t))\Big), \\
\alpha_2(t) & = &\varepsilon_2\Big( \cosh(\xi(t))+c_1\sin(K t+c_2)\sinh(\xi(t))\Big), \\
\alpha_3(t) & = &c_1\cos(K t+c_2).
\end{eqnarray*}
Finally, the mean curvature vector of $\Sigma_{\alpha}$ is
$$  \mathbf{H} = \frac{\cosh(\xi(t))\Big(1+c_1^2K^2\cos(2(K t+c_2))\Big)}{2c_1\cos(K t+c_2)\sqrt{1+(\alpha_{1}'(t))^{2}}}
(\varepsilon_1\eta_1-\varepsilon_2\eta_2).
$$
\end{example}

\begin{example}\label{ej:3} Given $K>0$, we compute the profile curve $\alpha$ of a surface of type B with constant
Gaussian curvature $-K^2$. By (\ref{eq:gauss1}), we need to solve
the differential equation $\alpha_3''(t)=K^2\alpha_3(t)$, whose
general solution is
$$ \alpha_3(t) = c_1\exp(K t)+c_2\exp(-K t), \quad \textrm{with} \ c_1,c_2\in\mathbb{R}, \ c_1^2+c_2^2>0.
$$
We choose an interval $I$ where $\alpha_3(t)$ is positive. We
discuss some cases.
\begin{description}
\item[{\em Case $c_1c_2> 0$}.] Given $\xi_0\in\mathbb{R}$, the angle function is
$$ \xi(t)\ =\ \frac{1}{K\sqrt{c_1c_2}} \arctan\left(\frac{c_1\exp(K t)}{\sqrt{c_1c_2}} \right)+\xi_0.$$
\item[{\em Case $c_1c_2< 0$}.] Given $\xi_0\in\mathbb{R}$, the angle function is
$$\xi(t)\ = \ \frac{1}{2K\sqrt{-c_1c_2}}\ln\left\vert \frac{2c_1\exp(K t)-2\sqrt{-c_1c_2}}{2c_1\exp(K t)+2\sqrt{-c_1c_2}}\right\vert\xi_0.
$$
\item[{\em Case $c_2=0$}.] Given $\xi_0\in\mathbb{R}$, the angle function is
$$ \xi(t) \ = \ -\frac{1}{K c_1 \exp(K t)}+\xi_0.$$
\item[{\em Case $c_1=0$}.] Given $\xi_0\in\mathbb{R}$, the angle function is
$$ \xi(t) \ = \ \frac{\exp(K t)}{K c_2}+\xi_0.$$
\end{description}
\noindent It only remains to compute $\alpha_1(t)$ and
$\alpha_2(t)$. To do so, we choose $\varepsilon_1$,
$\varepsilon_2=\pm 1$, and then
\begin{eqnarray*}
\alpha_1(t) & = &\varepsilon_1\Big( \sinh(\xi(t))-K\Big(c_1\exp(\xi(t))-c_2\exp(\xi(t))\Big) \cosh(\xi(t)) \Big), \\
\alpha_2(t) & = &\varepsilon_2\Big(
\cosh(\xi(t))-K\Big(c_1\exp(\xi(t))-c_2\exp(\xi(t))\Big)
\sinh(\xi(t)) \Big).
\end{eqnarray*}
Finally, the mean curvature vector of $\Sigma_{\alpha}$ is
$$ \mathbf{H}  \ = \ \frac{\cosh(\xi(t))
\Big(1-2K^2(c_1^2\exp(2Kt)+c_2^2\exp(-2K t))\Big)
}{2\alpha_3(t)\sqrt{1+\alpha_1'(t)^2)}}
(\varepsilon_1\eta_1-\varepsilon_2\eta_2).
$$
\end{example}

\begin{corollary} Let $S$ be a spatial surface in $\Lfour$ invariant by $\mathbf{G}_s$ satisfying $\|\mathbf{H}\|=0$ with constant Gaussian curvature $K$. Then, $S$ is locally congruent to either a surface of type A or one among Examples \ref{ej:1}, \ref{ej:2} and \ref{ej:3}.
\end{corollary}


\section{Conclusions}

In this paper, we have studied spacelike surfaces in Minkowski
4-space which are invariant by a rotation group of isometries and
whose mean curvature vector field is lightlike or zero. Our main
result is the classification of such surfaces in Theorem
\ref{th:class1}, from which it follows that there are two types of
surfaces, that we call of type A and B, which are not excluding.
As a consequence, a long list of corollaries is exhibited. Among
them, we locally describe MOTS and marginally trapped surfaces.
Furthermore, given up to countable infinitely many surfaces of
either type A or B whose mean curvature vector might be either
future or past-pointing, (and some reasonable conditions), we
describe a method to glue them in just one surface whose mean
curvature vector is null, which are invariant by a spacelike
rotation group, and having intermediate extremal regions among two
regions of type A or B. Also, we pay attention to the Gaussian
curvature, showing the possibility to construct surfaces of type B
with prescribed Gaussian curvature (at least, theoretically).
Among them, the list of surfaces with constant Gaussian curvature
is exhibited.

These constructions may lead to the study of generalized horizons
in Minkowski 4-space as well as in other spacetimes, since they
are foliated by marginally trapped surfaces.


\bibliographystyle{unsrt}

\end{document}